\documentclass[conference]{IEEEtran}

\usepackage{cite}

\usepackage{graphicx}
  \graphicspath{{./figures/}}
  \DeclareGraphicsExtensions{.pdf,.jpeg,.png,.eps}

\ifCLASSINFOpdf
\else
\fi

\usepackage[cmex10]{amsmath}
\usepackage{amssymb}

\begin{document}
%
\title{GPU Kernels for High-Speed 4-Bit Astrophysical Data Processing}

\author{\IEEEauthorblockN{
Peter Klages\IEEEauthorrefmark{1},
Kevin Bandura\IEEEauthorrefmark{2}, 
Nolan Denman\IEEEauthorrefmark{1},
Andre Recnik\IEEEauthorrefmark{1},
Jonathan Sievers\IEEEauthorrefmark{3},
and
Keith Vanderlinde\IEEEauthorrefmark{1},\IEEEauthorrefmark{4}}
\IEEEauthorblockA{\IEEEauthorrefmark{1}Dunlap Institute for Astronomy and Astrophysics, University of Toronto, Toronto, ON, Canada}
\IEEEauthorblockA{\IEEEauthorrefmark{2}Department of Physics, McGill University, Montr\'{e}al, QC, Canada}
\IEEEauthorblockA{\IEEEauthorrefmark{3}Astrophysics and Cosmology Research Unit, University of KwaZulu-Natal, Durban, South Africa}
\IEEEauthorblockA{\IEEEauthorrefmark{4}Department of Astronomy and Astrophysics, University of Toronto, Toronto, ON, Canada}
Contact Email: klages@dunlap.utoronto.ca
}

\maketitle

\begin{abstract}
	Interferometric radio telescopes often rely on computationally expensive $O(N^2)$ correlation calculations; fortunately these computations map well to massively parallel accelerators such as low-cost GPUs.  This paper describes the OpenCL kernels developed for the GPU based X-engine of a new hybrid FX correlator. Channelized data from the F-engine is supplied to the GPUs as 4-bit, offset-encoded real and imaginary integers.  Because of the low bit width of the data, two values may be packed into a 32-bit register, allowing multiplication and addition of more than one value with a single fused multiply-add instruction.  With this data and calculation packing scheme, as many as 5.6 effective tera-operations per second (TOPS) can be executed on a 4.3 TOPS GPU.  The kernel design allows correlations to scale to large numbers of input elements, limited only by maximum buffer sizes on the GPU.  This code is currently working on-sky with the CHIME Pathfinder Correlator in BC, Canada. 
\end{abstract}


%
\IEEEpeerreviewmaketitle

\section{Introduction}
The rapid development of Graphics Processing Units (GPUs), their ability to perform parallel computations on multiple data items,
and their low cost, driven by a very large consumer gaming market, 
has prompted researchers to develop techniques to utilize GPUs as general purpose computational accelerators for a variety of numerically intensive calculations.
For a certain subclass of programming problems, especially matrix methods, GPUs excel and can perform at several Tera FLoating point Operations Per Second (TFLOPS) per graphics accelerator.
In astrophysical research, the correlation of signals from elements in an interferometer is one such matrix method that is especially suited to GPU processing, and as such has seen development of solutions which take advantage of the easily accessible computational power of GPUs \cite{  2015arXiv150105992O, Clark:2013:ARA:2493921.2493925, HarrisEtAl, VanDerSchaaf}. 

FX correlators channelize the incoming data stream of each feed into many frequency bands (the F-engine), then correlate all $N$ of the incoming signals against all other signals for all the frequency bands (a cross correlation, hence X-engine).
FX correlators allow easy division of the $O(N^2)$ problem over multiple devices, as each frequency band can be processed independently.

The most popular and powerful GPU-based X-engine for FX correlators to date has been \texttt{xGPU}; this open-source code has been highly optimized over the years and achieves 
an impressive 91\% of the GPU's peak performance, with 79\% going directly to the pairwise correlation operations \cite{Clark:2013:ARA:2493921.2493925}.
However, since \texttt{xGPU} is programmed in NVIDIA's CUDA language, this limits hardware choices to NVIDIA GPUs.
	
To open X-engine code to more accelerator vendors and potentially more accelerator types,
a new high-efficiency, low-bit-depth X-engine correlator has been created\footnote{The OpenCL kernels are available under the MIT License at https://github.com/radiocosmology/CHIME-x} in OpenCL, a C-like language, for the Canadian Hydrogen Intensity Mapping Experiment (CHIME). 

While OpenCL itself is platform and vendor-nonspecific (any code that holds to OpenCL specifications will work on any accelerator that supports that specification), the optimizations described in this paper target AMD GPUs, specifically the Southern Islands family.
This $O(N^2)$ correlation algorithm can calculate multiple frequency bands simultaneously, scales well to large $N$ (currently tested to $N=8192$), and
is currently in use on the CHIME Pathfinder, a testbed and proving ground for the full-scale CHIME instrument. 
CHIME is a new, large-$N$ interferometric telescope under construction in Penticton, BC, Canada and
the Pathfinder is composed of 2 cylinders with 64 dual polarized feeds per cylinder, or $N=256$ total interferometric elements
\cite{2014SPIE.9145E..22B}.

The breakdown of the paper is as follows:
In \S\ref{sec:n2corr}, a basic correlator X-engine will be discussed, yielding metrics to compare the optimized kernel against, while
\S\ref{sec:n2alg} discusses the new 4-bit correlator.
In \S\ref{sec:discussion}, a discussion of extensions for these kernels will be given, along with their performances compared to their optimal theoretical performances.        
  
\section{$O(N^2)$ Correlation}
\label{sec:n2corr}

\subsection{Quantifying the Ideal Case}
Correlations are performed on channelized data received from the F-engine, with each frequency sub-band's data processed independently.
The correlation between two complex signals at one instant in time is:
\begin{align}\label{eq:correlation}
	\text{Corr}({X,Y}) ={}&{X} {Y}^* \nonumber \\
	={}&(X_{\text{Re}},X_{\text{Im}})(Y_{\text{Re}}, -Y_{\text{Im}}) \nonumber \\								
	={}& ((X_{\text{Re}}Y_{\text{Re}} + X_{\text{Im}}Y_{\text{Im}}), \nonumber \\ 									
	& (X_{\text{Im}}Y_{\text{Re}} - X_{\text{Re}}Y_{\text{Im}})	)	
\end{align}
where $X$ and $Y$ represent the input elements being correlated, and ${Y}^*$ refers to the complex conjugate of the input element data at index $Y$. Storing the real and imaginary components independently, there are 4 unique multiplies, 1 addition, and 1 subtraction per time step.
The correlation matrix, $A$, is the square matrix comprised of all $(X,Y)$ correlations, an outer product of the input stream data vector and its complex conjugate.
These products are accumulated over many time samples (typically thousands to millions) before the signal evolves
due to changes in telescope pointing or time variability in the sky.
Including the accumulation, a total of 8 arithmetic operations, or 4 fused Multiply-Add (MAD) instructions, are required per correlation matrix entry per time step.  
This set of 8 operations is often referred to as a complex Multiply-Accumulate (cMAC).

Since the correlation matrix, $A$, is Hermitian (i.e., $A_{ij}=A_{ji}^*$), only the upper triangle needs to be calculated,
so the total number of operations required per correlation matrix, per frequency band, per time step is:
	\begin{align}\label{eq:operationsIdeal}
		\text{Upper\_Triangle\_Ops}_{\text{Ideal}}=\frac{N\times(N+1)}{2}\times 8
	\end{align}
	where $N$ is the number of input elements.

\subsection{GPU Architecture Overview}
A GPU is composed of multiple Compute Units (CUs) that are each comprised of multiple stream processors (or compute cores).
Programs designed to run on the parallel compute cores are called kernels, and they describe what each work item in a work group does based on its local and global address.  
These work items, for AMD GPUs, will execute instructions on a single compute unit in lock step (or a pipelined manner if more work items than compute cores are grouped together), and have the ability to share local memory with each other, up to 32 kB.

Details about particular GPU architectures are readily available from their respective vendors, and various optimization strategies can be found in numerous books regarding OpenCL.
In general, to optimize GPU code (irrespective of the vendor), one must try to minimize accesses to slow memory types and hide the access latencies with many active threads and compute intensive kernels \cite{Volkov} while fitting the problem (or subsections of it) to the number of compute cores in a compute unit.

\subsection{Tiling}
\begin{figure}[!t]
\centering
\includegraphics[width=2.5in,bb=40 00 440 315]{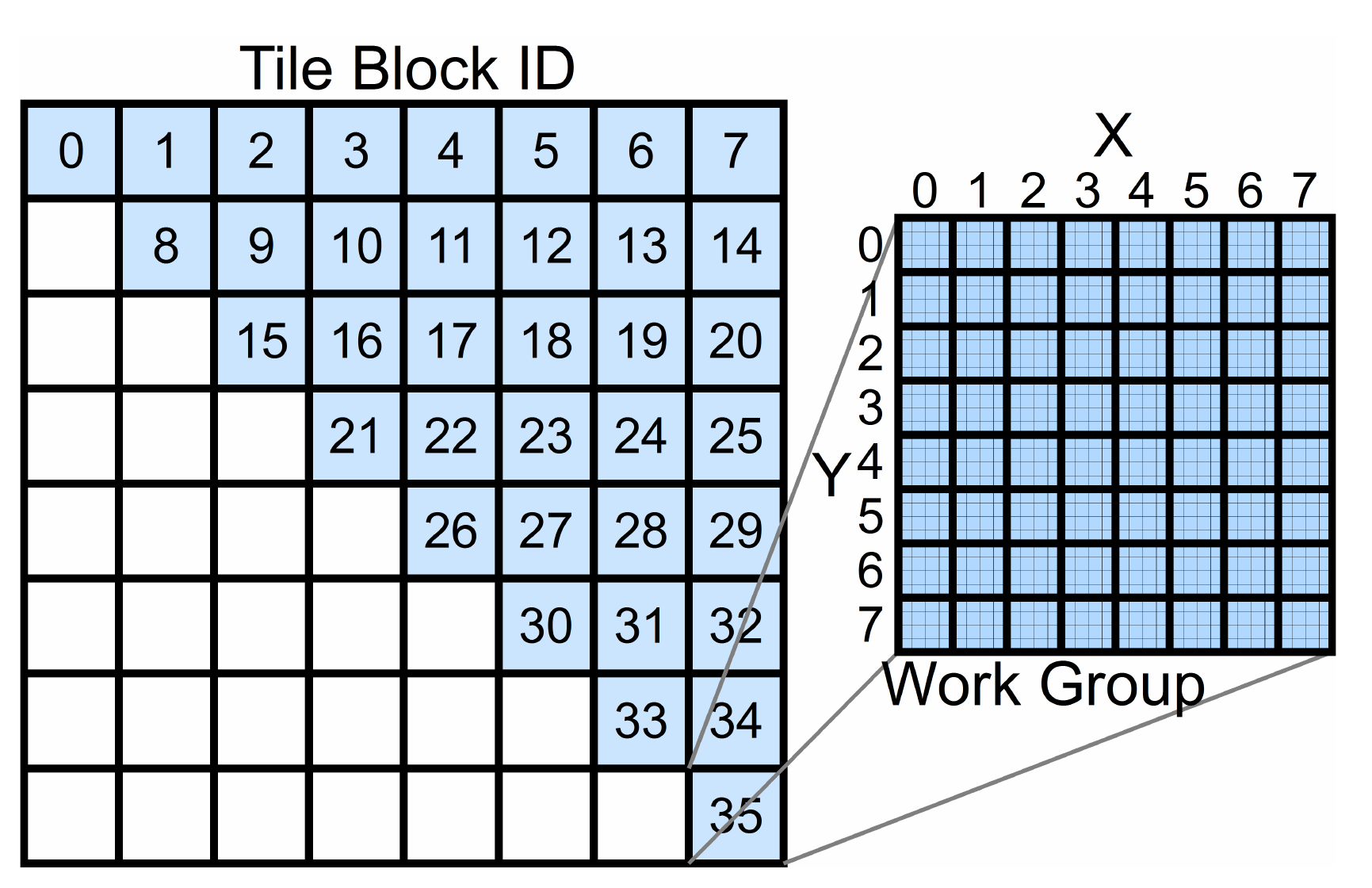}
\caption{Calculating the upper triangle of the correlation matrix using multiple square tiles of a smaller size.  In this figure, the upper triangle of a $256 \times 256$ matrix is tiled using 36 tiles of dimension $32 \times 32$.  Each of the tiles can be processed independently of all other tiles, and each tile is calculated using a 2D local work space ($8 \times 8$ work items).  Each work item calculates $4 \times 4$ complex correlation products.  }
\label{fig:tiling}
\end{figure}

Due to resource limitations, and to parallelize computations, the upper triangle of the large correlation matrix is divided into small blocks, or tiles, that can be computed independently  (Fig. \ref{fig:tiling}).
Each work group is responsible for calculating the entries in one tile, loading each element in the tile once from global memory, and sharing the value amongst the individual work items using higher speed local memory.

Tiling over the upper triangle of the correlation matrix
minimizes the number of redundant calculations and allows the number of loads from global memory to be kept relatively small.
Only the tiles that straddle the diagonal will have redundant calculations,
and, for large $N$, it is often more efficient to calculate and ignore the redundant results.
While larger tiles generally reduce the number of global memory reads, 
resource limitations and the cost of redundant on-diagonal calculations favor smaller tiles.
A tile size of $32 \times 32$ elements yields optimal performance 
for the algorithm described below, on AMD R9 280X GPUs, for large $N$ ($N \gtrsim 128$). 

\section{Optimized 4-bit Correlator}
\label{sec:n2alg}
\begin{figure}[!t]
\centering
\includegraphics[width=2.5in,bb=10 00 280 230]{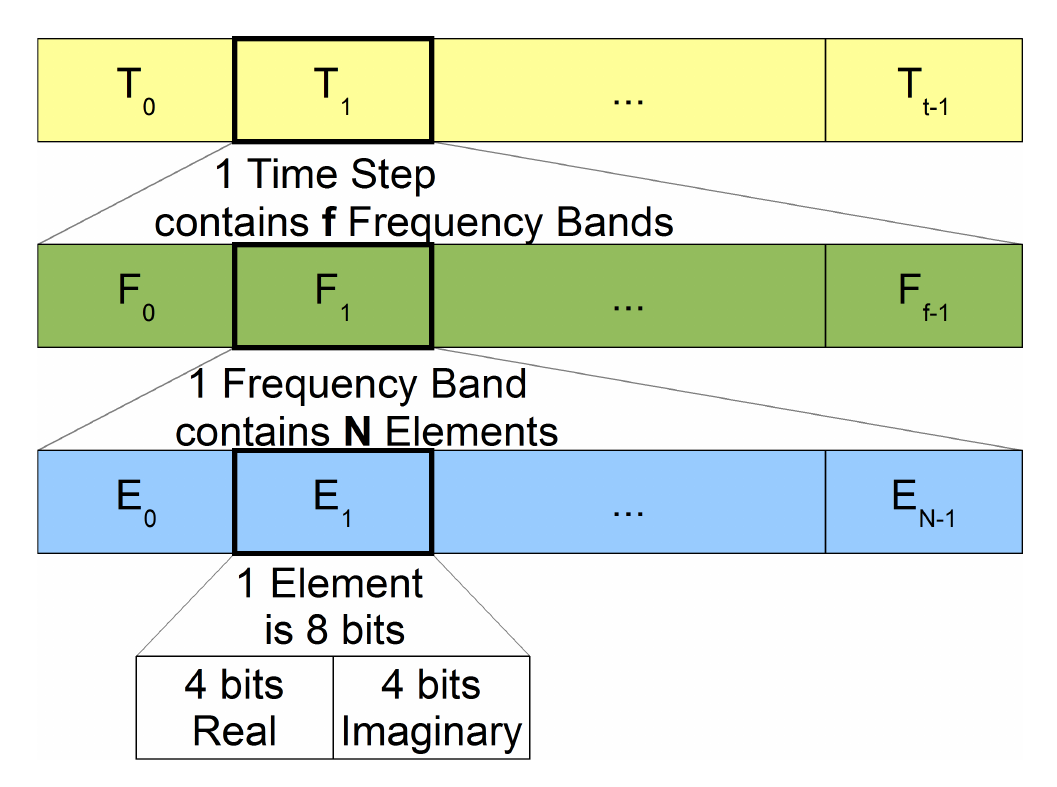}
\caption{The CHIME data stream is composed of multiple time steps, each of which are composed of multiple frequency sub-bands from the channelization process, which are in turn composed of $N$ input elements, each of which are 4+4-bit offset-encoded complex data.}
\label{fig:dataDescription}
\end{figure}

Input data arrives from an F-engine, represented
as offset-encoded unsigned 4-bit integers (also called Excess-8 encoding): data with a possible range $[-8,7]$ is shifted by an offset of 8 so that the numbers are represented as values in the range $[0,15]$. Both the real and imaginary values for one element are packed in a single byte (8 bits).
This packing arrangement for the input data is necessary in transmitting data from the FPGAs to the GPUs efficiently,
but does result in overhead related to unpacking the data to another variable type for processing.
The data for the individual elements are arranged in a consistent order for each frequency band, which repeats for each time step,
as illustrated in Fig. \ref{fig:dataDescription}.

Data packing techniques can also be used to leverage the low-bit-depth of the inputs to reduce the number of instructions required in a kernel.
Two 4-bit input values may be packed into a 32-bit register,
creating an implicit 2-tuple of \texttt{short integers} (16-bits) inside a 32-bit register.
When the 2-tuple is multiplied by a 4-bit number, two products are calculated simultaneously. 
By using the `\texttt{mad24}' function (a fused multiply-add which first multiplies two 24-bit numbers together then adds the lower 32-bit portion of the result to another 32-bit integer), a running sum of two products can be calculated in one clock cycle (see Fig. \ref{fig:unpack_and_packedMad24}b).
This packing effectively halves the number of critical instructions that are required, though it does impose a limit on the integration period in the kernel.
Since the maximum product of two 4-bit numbers is 225, a 16-bit storage container allows up to 291 time steps of products to be accumulated before overflows become possible.

The use of offset-encoded values means that additional terms must be calculated to undo the offsets introduced to the accumulated correlation matrix.
These offsets are:
\begin{align} \label{eq:offset}
  \text{B}{(X,Y,t)} = \sum_{Time=0}^{t}&{}{(- 8((X_{\text{Re}}+X_{\text{Im}})+(Y_{\text{Re}}+Y_{\text{Im}}))}
	+ 128, \nonumber \\&{}\text{ }
  8((X_{\text{Re}}-X_{\text{Im}})-
	(Y_{\text{Re}}-Y_{\text{Im}})))
\end{align}
where $B$ denotes the offset, $X$ and $Y$ refer to two offset-encoded complex input elements, and $t$ is the number of time steps being accumulated.
Undoing the systematic offsets for the output correlation products is only dependent on the running sums of the values of each input element,
with no interdependence between elements.
The offsets required for the final correlation output can be calculated in a combination of two kernels:
one kernel that accumulates the running sums and differences of the real and imaginary components
and one that combines the running sums and differences for a given number of time steps, $t$, as per Eq. \ref{eq:offset} to
pre-seed the correlation matrix entries with the correct offsets.
These additional kernels have minimal impact on the total computation time when $t \gg N$ (2\% or less).

\subsection{Optimized 4-bit $O(N^2)$ Kernel Description}
\begin{figure}[!t]
\centering
\includegraphics[width=2.5in, bb=50 10 540 760]{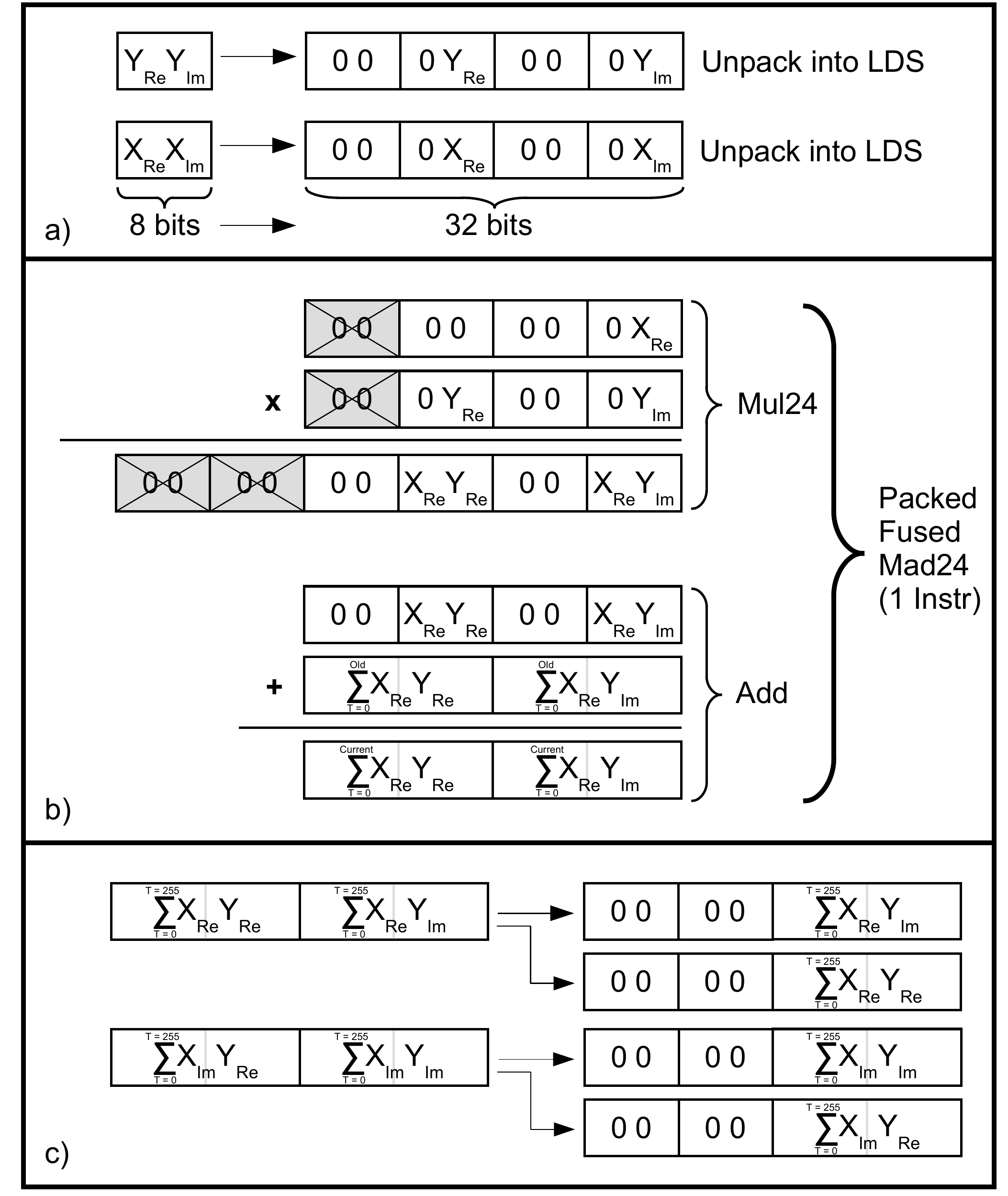}
\caption{
	a) Initial unpacking of 4+4-bit complex data in global memory to 16+16-bit complex data in shared local memory.  Unpacking and sharing local (LDS) memory reduces accesses to global memory and streamlines the critical code section.
	b) Multiplying two 4-bit numbers by another 4-bit number and accumulating into a 32-bit register can be done in one clock tick using the fused \texttt{mad24} instruction. The size of the products means that up to 291 multiplication and accumulations can be done in the packed 32-bit \texttt{unsigned integer}.
	c) The compressed data must be unpacked and converted to \texttt{signed integer} data before the results are able to be combined as in Eq. \ref{eq:correlation} and added to the matrix pre-seeded with offsets from Eq. \ref{eq:offset}.
}
\label{fig:unpack_and_packedMad24}
\end{figure}

The optimized 4-bit $O(N^2)$ correlation kernel operates as follows: 
\begin{itemize}
	\item Each work group contains $8 \times 8$ work items and corresponds to a tile on the correlation matrix.
	Each work item is responsible for calculating $4 \times 4$ complex correlation products,
	so each work group calculates a total of $32 \times 32$ complex correlation products.
	See Fig. \ref{fig:tiling}.
	\item The block ID for the tile is used to find offsets for the input element addresses for both $(X,Y)$ input elements to be correlated.
	 Additional offsets in the input buffer, based on the current time step, frequency band being processed, and work item ID are applied to the $(X, Y)$ input addresses. 
	\item There is a doubly nested loop which calculates and accumulates the correlation products over 256 time steps:
	\item In the outer loop, each work item loads and pre-expands the 4+4-bit complex data into 16+16-bit data using 32-bit \texttt{unsigned integers} in shared local memory (see Fig. \ref{fig:unpack_and_packedMad24}a).
	The shared local memory is used for high speed access in the computationally intense inner loop.
	Each iteration caches 8 time steps for the $X$ and $Y$ addresses to be computed in the inner loop.  
	\item In the inner loop, each work item loads the 4$\times X$ and 4$\times Y$ input elements for a single time step from the shared local data.
	The $X$ elements are divided into their real and imaginary components, and using `\texttt{mad24}' instructions, these are multiplied by the packed real and imaginary components of the 4 Y input elements.
	Only 4 clock ticks are required to multiply and accumulate 8 products because of the low-bit-width packed data and the fused \texttt{mad24} instruction (see Fig. \ref{fig:unpack_and_packedMad24}b).
	\item The outer and inner loops repeat, loading 8 more time steps during each outer loop iteration,
	processing and accumulating the products until 256 time steps have been accumulated.
	\item Following this accumulation, the output address for the block of correlation products is calculated.
	The output block is locked, and the accumulated products are extracted  (see Fig. \ref{fig:unpack_and_packedMad24}c),
	then added along with the previous results to the pre-seeded offset corrections (Eq \ref{eq:offset}),
	resulting in the correct, signed 32+32-bit complex correlation products.
	Finally, the memory block is unlocked and the kernel finishes.
\end{itemize}

\section{Results and Discussion}
\label{sec:discussion}
\begin{figure}[!t]
\centering
\includegraphics[width=3.6in, bb=0 20 450 500]{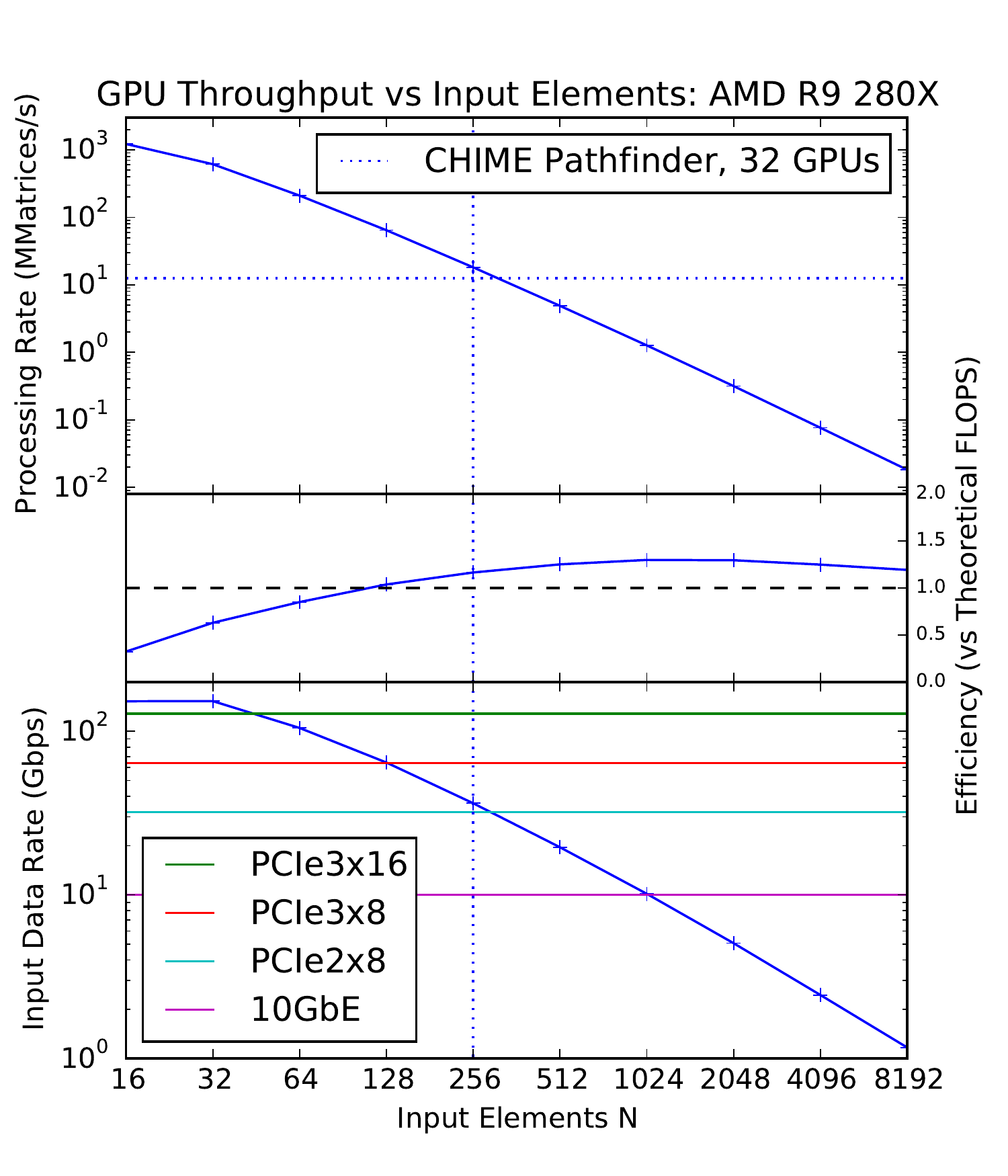}
\caption{
{Upper Panel:} Bandwidth processed, in millions of matrices/s, vs number of input elements, $N$. The number of correlation matrices calculated per second for a given $N$ shows the bandwidth that can be processed by a single AMD R9 280X GPU. To estimate the number of GPUs needed for a specific bandwidth, divide the desired bandwidth by the number of correlation matrices calculated per second. For full CHIME, $N = 2048$, BW$_{\text{desired}}$ = 400\,MHz, and the tested GPU computation rate (331800 correlation matrices/s; refer to the plot), $~\sim$ 1200 GPUs would be needed for the full $O(N^2)$ correlation. Middle Panel: Packed algorithmic compute efficiency divided by the unpacked theoretical efficiency. Lower Panel: I/O Bandwidth required for each GPU to compute at the rates shown in the Upper Panel. For CHIME Pathfinder, PCIe 3 is required for optimal I/O, however as $N$ increases above $N$ = 256, the I/O requirements become quite low and only PCIe2 is needed to transfer data to the GPU.}
\label{fig:BWperformance}
\end{figure}

The algorithm achieves high efficiency over a broad range of $N$, peaking at $N=2048$ inputs, corresponding to the full CHIME instrument (see Fig. \ref{fig:BWperformance}b).
On an AMD R9 280X, the nominal maximum computation rate is 4.3 TOPS (single precision floating point numbers or integers), using fused MAD instructions\footnote{1050 MHz $\times 32$ compute units $\times 64$ cores $\times 2$ ops/MAD = 4.3 TOPS}.
For $N=2048$, this corresponds to a maximum of 256 thousand correlation matrices per GPU per second (using Eq. \ref{eq:operationsIdeal}).
Using the new, optimized 4-bit correlator with packed arithmetic instructions, 336 thousand correlation matrices can be calculated per second,
or 131\,\% of the maximum that would be possible without packing the data.  
The packed arithmetic implementation described here does not lead to a full doubling in efficiency because in the critical code section an additional unpack operation per 4 \texttt{mad24} operations is required.  This extra operation limits the benefits from this packing scheme to $1.6\times$, 
meaning this implementation runs at $\sim$ 82\,\% of the peak calculation rate of the GPU.
Improving this algorithm significantly would require bypassing the unpacking operation; this could be done with a \texttt{mad24\_hi} instruction
(accumulating the overflow 16 bits from a \texttt{mul24}), but no such instruction is included in the OpenCL 1.2 specification nor AMD-specific extensions.
While the kernel optimizations are dependent on both the GPU model and the driver being used, once a card and driver are set, the parameters can be optimized relatively easily.

Fig. \ref{fig:BWperformance} shows the performance of the algorithm on AMD R9 280X GPUs, plotting
the number of correlation matrices processed per second vs $N$ in the top panel, a comparison to the ideal method (without data packing) in the middle panel, and the implied input data rates in the bottom panel.
The algorithm scales well to large-$N$, limited only by buffer limitations on the output data. Changing the number of frequency sub-bands processed simultaneously in a kernel has a minimal effect on processing efficiency. 
This figure also allows one to estimate the number of GPUs that would be required for a given instrument with current GPU technology: 
the 400 MHz bandwidth of the CHIME Pathfinder can easily be processed by 32$\times$ R9 280X GPUs.

While the algorithm presented here allows one to correlate multiple frequency bands simultaneously for a scalable number of input elements, extensions to the code base have also been created. 
A kernel has been developed that modifies the input stream at the raw data level, allowing individual elements to be shifted by integer time steps pre-correlation; lags can be generated or compensated for computationally, which is useful with varying cable lengths.
The correlation kernel has also been extended to gate the correlation products on multiple time periods; gating the products is useful for analysis of data where periodic signals occur (e.g., artificial calibration signals, and pulsars). 

\section{Conclusion}
A 4-bit X-engine for a heterogeneous FX correlator has been created that computes correlation matrices on GPUs at a higher rate than the peak floating point (or integer) operation rate natively allows.  It does this by packing two 4-bit numbers into a 32-bit register and taking advantage of the fused \texttt{mad24} instruction on the GPU. 
Comparing this packed, tiled algorithm to the ideal upper triangle correlation method at peak GPU computation rates, this algorithm executes at 131\,\% of the maximum non-packing performance ($N=2048$). While there is some computational overhead associated with the packing, and the total hardware usage of this implementation is lower than \texttt{xGPU}'s (81\,\% vs 91\,\% \cite{Clark:2013:ARA:2493921.2493925}), this packing scheme outperforms the CUDA implementation on 4-bit inputs (131\,\% vs 79\,\% \cite{Clark:2013:ARA:2493921.2493925} peak performance).  For any new correlator to outperform this method, packed operations of some kind must be employed.

The code is open source, and while optimized for AMD Southern Islands GPUs, it was written in the vendor-nonspecific language, OpenCL; re-optimizing this code for other platforms with similar abilities should not be difficult.

\section*{Acknowledgments}

Peter Klages thanks IBM Canada for funding his research and work through the Southern Ontario Smart Computing Innovation Platform (SOSCIP).
We are very grateful for the warm reception and skillful help we have received from the staff of the Dominion Radio Astrophysical Observatory, which is operated by the National Research Council of Canada.
We acknowledge support from the Canada Foundation for Innovation, 
the Natural Sciences and Engineering Research Council of Canada, 
the B.C. Knowledge Development Fund,  
le Cofinancement gouvernement du Qu\'ebec-FCI, 
the Ontario Research Fund, 
the CIfAR Cosmology and Gravity program, 
the Canada Research Chairs program, and 
the National Research Council of Canada.
We thank Xilinx University Programs for their generous support of the CHIME project,
and AMD for donation of test units.



\bibliographystyle{IEEEtran}
%
\bibliography{kernelPaper}

\end{document}